\documentclass{aastex}
\usepackage{emulateapj5,graphicx}

\begin{document}

\newcommand{\bm}[1]{\mbox{\boldmath $#1$}}
\title{
Photodissociation feedback of Population III stars \\on their neighbor prestellar cores
}
\author{Hajime Susa\altaffilmark{1} 
\vskip 0.2cm
\affil{Department of Physics, Rikkyo University, Nishi-Ikebukuro,
Toshimaku, Tokyo, Japan}
\altaffiltext{1}{susa@rikkyo.ac.jp}
}
\begin{abstract}
We investigate the star formation process in primordial environment in
 the presence of radiative feedback by other population III stars formed
 earlier. In this paper, we focus our attention on the effects by photodissociative
 radiation toward the full understanding of the radiative feedback effects.
 We perform three dimensional radiation hydrodynamics simulations on this
 issue as well as analytic estimates, paying special attention on the
 self-shielding effect and dynamics of the star-forming cloud. 
  As a result, we
 find that the ignition timing of the source star is
 crucial. If the ignition is later than the epoch when the central
 density of the collapsing cloud exceeds  $\sim 10^3-10^4{\rm cm^{-3}}$,
 the collapse cannot be reverted, even if the source star is located
 at $\la$ 100pc. The uncertainty of the critical
 density comes from the variety of initial conditions of the collapsing cloud.
We also find the analytic criteria for a cloud to collapse with given 
central density, temperature and the Lyman-Werner(LW) band flux which
 irradiates the cloud. 
 Although we focus on the radiation from
 neighbor stars, this result can also be applied to the effects of
 diffuse LW radiation field, that is expected to be built up prior to
 the reionization of the universe. 
We find that self-gravitating clouds can easily self-shield from 
   diffuse LW radiation and continue their collapse for densities
   larger than $\sim 10^3 {\rm cm^{-3}}$.

\end{abstract}
\keywords{theory:early universe --- galaxies: formation --- radiative transfer 
--- molecular processes --- hydrodynamics}

\section{Introduction}
\label{intro}
One of the important objectives of cosmology today is to understand 
the way how the first generation stars or galaxies are formed and how they
affected the later structure formation, and how they reionize the
surrounding media. 
In particular, the self-regulation of star formation in the first low
mass halos is quite important, since it controls the Population III
(POPIII) star formation
activity in the early universe, which is the key for the reionization
and the metal pollution of the universe.
These problems have been studied intensively in the last decade, hence we
already have some knowledge on these issues (see review by Bromm \&
Larson 2004). 
However, studies which
properly address the radiation transfer effects are still at the
beginning, in spite of their great importance.

Radiation from the first generation stars play quite important roles on the
formation of stars/galaxies through two main physical processes. Since
first stars are expected to be very massive \citep{ON98,Abel00,BCL02,NU99,NU01}, the
amount of ultraviolet photons are quite large. Once these kinds of stars
are formed, 1) emitted Lyman-Werner (LW) band photons photodissociate the
hydrogen molecules in the neighborhood by Solomon
process\citep{HRL97,ON99,GB01,Macha01}. 
Afterwards or
simultaneously 2) ionizing photons propagate into surrounding media,
followed by the
photoionization and heating of the gas. Since hydrogen molecules are the only
coolant in low mass first generation objects, and the photoheated gas is
too hot to be kept in such small dark halos, these radiative feedbacks
are basically expected to have negative effects on the structure
formation (e.g. Susa \& Umemura 2004ab).
In contrast, enhanced fraction of electrons by photoionization can lead
to efficient H$_2$ formation \citep{SK87, KS92, SUNY98, OhH02}, and it
could accelerate the structure formation \citep{Ricotti01, Oshea05,
Nagakura05, SU06, AS06, Yoshida06}.

Since these physical processes related to this issue are so complicated, 
we try to understand one by one. 
In this paper, we focus our attention on the photodissociation feedback
effect including the effect of self-shielding.
We do not take into account the effects by photoionization, which is discussed
with limited parameters in \citet{SU06}, and will be presented 
in detail in the forthcoming paper.

Concerning the photodissociation feedback effects, there are two
contexts. 
First one is the local feedback by the nearby POPIII stars
formed in the same halo or in the neighbor halos. 
Another one is the feedback by diffuse LW background
radiation, that is built up by the formed POPIII stars in many dark
halos spread in the universe\citep{Macha01,Yoshida03}. 
In reality, the former would particularly be
important at the early phase of the POPIII star formation episode,
ahead of the latter effect. Local feedback might also play important
roles in the star formation process in relatively larger halos whose
virial temperature is higher than $10^4$K.
The physical difference of the
two cases are the different duration and intensity of radiation field. 
In the latter case,  intensity is not so high, whereas the radiation
field irradiate the halos continuously since it is a background. 
In contrast, in the former case, the radiation source
basically turn-off after the life-time of the source star, although
the intensity of 
the radiation could be much higher than the diffuse background.
In this paper, we mainly investigate the local effects.

\citet{ON99} investigated on this issue in an uniform virialized
halo. They argue that only a single massive POPIII star can destroy H$_2$
molecules within $\sim 1$kpc, which is comparable to the virial radius
of $10^8 M_\odot$ halos at $z\ga 10$. This means that star formation in
such low mass halos is highly regulated by themselves. On the other
hand, \citet{GB01} considered more realistic clumpy halos. They
simply assessed the photodissociation time scale and free-fall time of
the dense clouds.
Comparing these two time scales, they
found that if the gas density is sufficiently high, dissociation time
scale could be longer than the free-fall time, which allow the cloud to
collapse. Although those arguments are quite clear, they are based upon very
simplified configuration. Therefore, more sophisticated treatments are
necessary to investigate this feedback process. In order to improve the
understanding of this issue, we
carry out three dimensional radiation hydrodynamics simulations as well as
more detailed analytic arguments.

This paper is organized as follows: in the next section, we briefly
summarize the numerical scheme employed. In \S\ref{setup}, setup of our
numerical simulations are shown. We present the numerical and
analytical results in \S\ref{resultLW} and \S\ref{analytic}. \S 6 and \S
7 are devoted to discussion and summary.

\section{Numerical method}
We perform numerical simulations by Radiation-SPH code developed by
ourselves \citep{Susa06}.
 In this section, we try to briefly summarize the scheme. 
The code is designed to investigate the
formation and evolution of the first generation objects at $z \ga 10$,
where the radiative feedback from various sources play important
roles. The code can compute the fraction of chemical species e, H$^+$, H,
H$^-$, H$_2$,  and H$_2^+$ by fully implicit time integration. It also
can deal with multiple sources of ionizing radiation, as well as
the radiation at Lyman-Werner band.

Hydrodynamics is calculated by Smoothed Particle Hydrodynamics (SPH)
method. We use the version of SPH by \citet{Ume93} with the modification
according to \citet{SM93}. We adopt the particle resizing
formalism by \citet{Thac00}.

The non-equilibrium chemistry and radiative cooling 
for primordial gas are calculated by the code
developed by \citet{SuKi00}, where H$_2$ cooling and
reaction rates are mostly taken from \citet{GP98}.
As for the photoionization process, 
we employ so-called on-the-spot approximation
\citep{Spitzer78}, but we do not get into the details since we do not
show the results with ionizing radiation in this paper.

The optical depth is integrated utilizing the 
neighbour lists of SPH particles. 
It is similar to the code described
in \citet{SU04a}, but now we can deal with multiple point
sources. In our new scheme, we do not create so many grid points on the light
ray as the previous code \citet{SU04a} does. 
Instead, we just create one grid point per one SPH particle in its
neighbor. We find the 'upstream' particle for each SPH particle on its
line of sight to the source. Then the optical depth from the source to
the SPH particle is obtained by summing up the optical depth at the 'upstream'
particle and the differential optical depth between the two particles.

The opacity of LW radiation is calculated by utilizing the original
self-shielding function\citep{DB96}. The column density of H$_2$ is
evaluated by the method described above. We assume same absorption for
the photons within LW band, i.e. between 11.26eV and 13.6eV.

The code is already parallelized with MPI library. The computational
domain is divided by so called Orthogonal Recursive Bisection method\citep{Dubinski96}. 
The parallelization method for radiation transfer part
is similar to Multiple Wave Front method developed by \citet{NUS01} and
\citet{Hienemann05}, but it is changed to fit the SPH scheme.
The code also is able to handle self-gravity with Barnes-Hut tree\citep{BH86}, which is
also parallelized.
The code is tested for various standard problems\citep{Susa06}. 
We also take part in the code comparison
project with other radiation hydrodynamics codes \citep{TSU3} in
which we find reasonable agreements with each other.

\section{Setup of Numerical simulations}
\label{setup}
We simulate the collapsing primordial gas cloud with $M = 8.3\times
10^4 M_\odot$, which is initially uniform and spherical. 
The initial number density is $14 {\rm cm^{-3}}$, whereas the
corresponding radius is $\sim 40$pc.
We employ four initial temperatures $T_{\rm ini}=50$K, $100$K
, $200$K and $350$K which represent the coldness of the initial collapses.
The uniform cloud is embedded in an extended uniform low density cloud
with $0.1 {\rm cm^{-3}}$ and the same temperature as the dense cloud.
Remark that such a thin surrounding material does not affect the fate of
the dense cloud.
The mass of an SPH particle is $0.32M_\odot$. We employ 524288 particles
in total. Approximetely half of the particles are used to represent the
initial dense region. 
The chemical
compositions are assumed to be the cosmological residual values \citep{GP98}. 

We start numerical simulations from this initial setup without any
radiative feedback effects. Then, the dense clump collapses in run-away
fashion because of the self-gravity and the radiative cooling by H$_2$ .
When the central density exceeds a certain limit, $n_{\rm on}$, we ignite a
120$M_\odot$ POPIII star that is located $D$ pc distant from the
center of the dense clump. The luminosity and the effective temperature
of the star are taken from \citet{Baraffe01}.
We employ various parameters in the range $3\times 10^2 {\rm cm^{-3}}\le
n_{\rm on} \le 1\times 10^5 {\rm cm^{-3}}$, and $20{\rm pc} \le D \le 200{\rm pc}$.
In this paper, the ionizing photons are not included, since we focus on the photodissociation process.
Simulated physical time is $10^7$ yr, which is significantly longer than 
the life time of a 120$M_\odot$ star ($\sim2$Myr). 
During the simulation
time, we do not turn-off the source star, as we are interested in
both of the effects by diffuse background/local LW radiation (diffuse radiation
never turns-off).

Numerical simulations are mainly carried out on parallel computer system called
``FIRST'' installed in University of Tsukuba. The newly being developed PC cluster
``FIRST'' is designed to elucidate the origin of first generation objects
in the Universe through large-scale simulations.
The cluster consists of 240 nodes 
( each node has dual Xeon processors ).
The most
striking feature of the system is that each node has small GRAPE6
PCI-X board called Brade-GRAPE. The Brade-GRAPE has four GRAPE6 chips
per board (thus 1/8 processors of a full board). 
The board has basically the same function as
GRAPE-6A \citep{FMK05}, but it has an advantage that it can be
installed in 2U server unit of PC clusters due to the short height of
its heat sink.

\section{Numerical Results}
\label{resultLW}
\subsection{Simulation without radiative feedback}
Before we proceed to the results with radiative feedback, we give an
outline of the results without the feedback process. 
Since we have no LW flux in these particular runs
 and total mass and the initial density are fixed, 
the only free parameter of these calculations is the 
initial temperature, $T_{\rm ini}$. Fig.\ref{fig:rhor_norad} shows the
snapshots of radial density distributions of the collapsing clouds with
$T_{\rm ini}=100$K and $T_{\rm ini}=350$K. The snapshots are taken when
the central dense regions are very close to the resolution limits. For
both of the runs, the distribution roughly converges to the $\rho \propto
r^{-2}$relation,  which is well known as the Larson-Penston type
similarity solution \citep{Larson69,Penst69} for isothermal gas. 
Remark that the larger $T_{\rm ini}$ results in 
smaller size of the cloud, since higher temperature promotes
faster reaction to produce H$_2$, which leads to efficient cooling.

Fig.\ref{fig:norad} represents the evolution of the cloud core on
density v.s. temperature plane. Four tracks correspond to $T_{\rm ini}
=50,100,200,350$K, respectively. Among these, the run with $T_{\rm ini}
=350$K is relatively close to the physical states of the collapsing
clouds at $n_{\rm N}\ga 10^3 {\rm cm^{\-3}}$, developed from the CDM
density fluctuations (see figures 11,12 in O'shea \& Norman 2006).
The runs with $T_{\rm ini} = 50, 100$K also could be realized in the
shock compressed layers in larger halos with $M\ga 10^{8-9}
M_\odot$ since the gas is cooled down to $\la 100$K due to the efficient
H$_2$ cooling\citep{SK87,KS92,SUNY98,UI00}.
Thus, the various $T_{\rm ini}$ should be regarded as the variety of the
initial conditions for the sites of primordial star formation.

\subsection{Typical results with radiative feedback}
First, we show two typical results among various runs. 
Figure \ref{fig:40pc} shows 
the density / H$_2$ map on the slice that
contains the axis of symmetry 
in the run with $T_{\rm ini}=100{\rm K}, ~n_{\rm on} =
10^3{\rm cm^{-3}}, D=40{\rm pc}$. 
Top panels represent three snapshots,
those correspond to  $t=10$yr, $10^4$yr and $10^6$yr, respectively. 
Bottom panels also show $y_{\rm H_2}$ distribution at same physical
time. In this run, H$_2$ are totally photodissociated by the nearby
star, even at the center of the collapsing cloud (lower panels).
H$_2$ fraction never recovers through the simulation time.
 As a
result, the cloud cannot keep collapsing because of the absence of
coolants. The cloud bounces after the adiabatic compression phase.

On the other hand, Figure \ref{fig:100pc} shows the case same as Figure
\ref{fig:40pc} except $D=100$pc. Since the Lyman-Werner band flux is
smaller than the previous case, H$_2$ is self-shielded and survive at
the center (lower panels),
followed by the collapse of the core (upper right panel).
Figure \ref{fig:h2center} shows the time
evolution of the central H$_2$ fraction for four models including above
two. The four models correspond to $D=40,60,80,100$pc respectively,
whereas other parameters $n_{\rm on}=10^3 {\rm cm^{-3}}$ and $T_{\rm
ini}=100{\rm K}$ are common.
The horizontal
axis shows the time after the collapse started, while the vertical axis
shows the H$_2$ fraction at the center of the cloud. We find that H$_2$
molecules are photodissociated just after the ignition of the source
stars for all models. After the dissociation,  
H$_2$ fraction recovers immediately. For the models with $D=80,100$pc,
the recoveries are 
sufficient to cool the gas ($y_{\rm H2}\sim 10^{-4}$ is
required), whereas it is not true for $D=40$pc case. $D=60$pc is the
marginal case, in which the cloud manage to collapse although it is delayed
significantly, because the cooling time is slightly longer than the
free-fall time. 
We will discuss
this delicate behaviour of H$_2$ fraction in
section \ref{h2fraction}.
\subsection{Summary of numerical runs}
Figure \ref{fig:all} shows the results of various runs. Four panels
correspond to $T_{\rm ini}=50,100,200,350$K, respectively. Horizontal
axes represents the turn-on density $n_{\rm on}$ and vertical axes show
the distance between the source star and the cloud center, $D$. Each
symbol in the panels corresponds to each run. In the runs denoted by
vertices, the clouds cannot collapse until the end of the simulation,
i.e. $10^7$yr after the ignition of the source star. 
On the other hand, the open circles represent the runs with
successful collapse. Here ``collapsed cloud'' is defined as the one
whose central density exceeds the numerical resolution limit determined
by the Jeans condition. The nominal density correspond to the
resolution limit is $\sim 10^6 - 10^7{\rm cm^{-3}}$ in the present paper.

The results depend upon the initial
temperature $T_{\rm ini}$ i.e. initial coldness of the collapsing
cloud. Initially colder gas can collapse for wider range of parameters
than those in hotter initial conditions. In fact, considering the
case with $T_{\rm ini}=350$K (lower left panel), which is close to the prestellar core
formed directly from CDM density perturbations, LW
feedback is inefficient for $n_{\rm N} \ga 10^4 {\rm
cm^{-3}}$ for $D\la 50{\rm pc}$. 
Thus LW feedback from the stars in
the same halo would not be effective for such dense self-gravitating cores,
since the virial radius of the $\sim 10^6 M_\odot$ halos at $z\sim
20-30$ is several ten parsec.

On the other hand, considering $T_{\rm ini}=50,100$K cases, LW feedback
is also ineffective (upper panels) even in less dense clouds ($n_{\rm N}
\ga 10^3 {\rm cm^{-3}}$). 
Thus, if we regard
these as the stars formed
from the shocked layers or fossil HII regions, such POPII.5
stars\citep{JB06,GB06} would not be affected by LW radiation even if the source stars are in the same halo.

These results are discussed
physically in detail in the next section.

\subsection{Effects of diffuse LW flux on dense cores}
 
Before we move on to the detailed physical arguments, we evaluate the effects of diffuse LW radiation field on the
primordial star formation. Previous analysis on this issue is basically
based upon the optically thin approximation (e.g. Machacek, Bryan \&
Abel 2001), which is only valid for low density clouds. \citet{Yoshida03} also address this
issue including the effects of self-shielding in their cosmological
simulations, although in an approximated manner. Present calculation can
give a better understanding on this issue especially for dense clumps.

The intensity of the diffuse LW radiation background formed after
the first generation stars, is expected to be $J_{\rm LW}\la 10^{-21}{\rm
erg/s/Hz/cm^2/str}$ for $z > 20$ \citep{CFA00,MBH06}. If we interpret this
intensity as the flux from distant 120$M_\odot$ single POPIII star, the distance $D$ should satisfy $D\ga 1{\rm kpc}$. 

In view of the numerical results in Figure \ref{fig:all}, diffuse
radiation field cannot prevent the self-gravitating prestellar cores
($n_{\rm N}\ga 10^3{\rm cm^{-3}}$) from collapsing, although the diffuse
field keep irradiating the cloud more than two million years.
However, it would be possible to stop the collapse at much lower
density, at which the clouds are not expected to be
self-gravitating (e.g. Ahn \& Shapiro 2006). In such phase, dark matter gravitational force dominates the
contraction, which is not included in the present paper. It is
beyond the scope of our present calculations.

\section{Analytic estimation}
\label{analytic}
In this section, we describe the analytic collapse criteria in the presence of
photodissociative feedback by a nearby star. 
The collapse criteria should
be $t_{\rm cool} < t_{\rm ff}$, i.e. the cloud is cooled faster than it
shrinks by gravity. After some algebra, an analytic expression for $t_{\rm cool} < t_{\rm
ff}$ is derived, which is compared with numerical results. We also
compare the derived cooling condition with $t_{\rm dis} > t_{\rm ff}$.
\subsection{Non-equilibrium fraction of electrons}
In order to assess the cooling time, we need the
number density of hydrogen molecules. The amount of hydrogen
molecules strongly depends on the electron abundance, since they are
mainly formed through following reactions:
\begin{eqnarray}
{\rm H} + {\rm e}^- &\rightarrow& {\rm H}^- + h\nu \label{eq:hminus}\\
{\rm H^-} + {\rm H} &\rightarrow& {\rm H}_2 + {\rm e}^-\nonumber
\end{eqnarray}  
Thus, we also have to assess the amount of electrons, which is out of
equilibrium.
The fraction of electrons ($y_{\rm e}$) at the center of the collapsing
cloud is not in chemical
equilibrium, which follows the following rate equation:
\begin{equation}
\frac{dy_{\rm e}}{dt} = -k_{\rm rec}n_{\rm N}y_{\rm e}^2.
\end{equation} 
Here the collisional ionization term is omitted, because the
temperature of the core is much lower than $10^4$ K above which the
collisional ionization becomes important. Since the central part of the
core collapses with free-fall time, the evolution of the hydrogen
nucleon number density $n_{\rm N}$ is described as,
\begin{equation}
\frac{1}{n_{\rm N}}\frac{dn_{\rm N}}{dt} \simeq \left(\frac{3\pi}{32Gm_{\rm p}n_{\rm N}}\right)^{-1/2}
\end{equation} 
Combining above two equations, we have
\begin{equation}
\frac{dy_{\rm e}}{dn_{\rm N}} \simeq -n_{\rm N}^{-1/2} y_{\rm e}^2 k_{\rm
 rec}\sqrt{\frac{3\pi}{32Gm_{\rm p}}} \label{eq:ye_den}.
\end{equation}
$k_{\rm rec}$ is a function of time, because it depends on
the temperature ($\propto T^{-1/2}$). 
However, the change of temperature is not so significant
during the collapse, whereas the density gets larger by several orders of
magnitude. Thus, we can approximate the recombination rate $k_{\rm rec}$ as
a constant when we integrate the equation (\ref{eq:ye_den}).
The equation (\ref{eq:ye_den}) has an analytic solution:
\begin{equation}
y_{\rm e} = \frac{1}
{y_{\rm e0}^{-1} + 2\left(\sqrt{n_{\rm N}}-\sqrt{n_{\rm N0}}\right)
k_{\rm rec}\sqrt{\frac{3\pi}{32Gm_{\rm p}}}}\label{eq:ye_exact},
\end{equation}
where $y_{\rm e0}$ denotes the initial
electron fraction, and $n_{\rm N0}$ denotes the initial number density
of hydrogen nucleus.
Consequently, we obtain the expression of $y_{\rm e}$ as a function of
number density $n_{\rm N}$. 
The asymptotic behaviour of the solution (\ref{eq:ye_exact}) for $n_{\rm
N}\gg n_{\rm N0}$ is 
\begin{equation}
y_{\rm e} \simeq \frac{1}{\sqrt{n_{\rm N}}} \frac{1}{k_{\rm rec}}\sqrt{\frac{32Gm_{\rm p}}{3\pi}}\label{eq:ye_app}.
\end{equation}

This expression is independent of initial abundance $y_{\rm e0}$ and
initial density $n_{\rm N0}$, except
that $y_{\rm e}$ cannot exceed $y_{\rm e0}$ since we neglect the
ionization processes. In fact, $y_{\rm e}$ given in equation
(\ref{eq:ye_exact}) are plotted for various
initial conditions $(n_{\rm N0}, y_{\rm e0})$ in Figure \ref{fig:ye}.
We find rapid convergences of electron fractions to equation
(\ref{eq:ye_app}) as the collapse of the clouds proceed. Therefore, as far as we consider
$n_{\rm N} \ga 10^3 {\rm cm^{-3}}$, we can safely assess the
electron fraction by equation (\ref{eq:ye_app}).

\subsection{H$_2$ abundance}
\label{h2fraction}
Because of the intense LW radiation, H$_2$ at the center of the cloud is
basically in chemical equilibrium with given $y_{\rm e}$.  Therefore we
assess the H$_2$ fraction $y_{\rm H_2}$ assuming chemical
equilibrium. Equating the formation rate of H$_2$ and the
photodissociation rate, we have
\begin{equation}
y_{\rm H_2} = \frac{n_{\rm N}y_{\rm e}k_{\rm H^-}}{k_{\rm 2step}} \label{eq:eqh2}
\end{equation}
where $k_{\rm H^-}$ denotes the reaction rate of reaction(\ref{eq:hminus}),
$k_{\rm 2step}$ is photodissociation rate by Solomon process. 
These rates are
\begin{eqnarray}
k_{\rm H^-} &=& 1.0\times 10^{-18} T {\rm cm^{3} s^{-1}}\label{eq:form}\\
k_{\rm 2step}&=& 1.13 \times 10^8 F_{\rm LW0}f_{\rm sh}\left(\frac{N_{\rm
						       H_2}}{10^{14}{\rm
						       cm^{-2}}}\right) {\rm s^{-1}},\label{eq:dis}
\end{eqnarray}
where $F_{\rm LW0}$ is the LW flux from the star in the absence of
shielding, $f_{\rm sh}$ denotes the self-shielding function derived by
\citet{DB96}. It is defined as
$$
f_{\rm sh}(x) = \left\{
\begin{array}{cc}
1,~~~~~~~~~~~~~~x \le 1 &\\
x^{-3/4},~~~~~~~~~x > 1 &
\end{array}
\right.
$$
$N_{\rm H_2}$ denotes the hydrogen column density of the collapsing
core. Since the core size is approximately given by the Jeans length,
$N_{\rm H_2}$ is given as
\begin{equation}
N_{\rm H_2}\simeq n_{\rm N} y_{\rm H_2} \frac{\lambda_{\rm J}}{2} = 
n_{\rm N} y_{\rm H_2}\frac{1}{2}\sqrt{\frac{\pi k_{\rm B} T}{G m_{\rm p}^2 n_{\rm N}}}\label{eq:Nh2}
\end{equation}
 
Combining equations (\ref{eq:ye_app})-(\ref{eq:Nh2}), we obtain
\begin{eqnarray}
y_{\rm H_2} = 4.3&\times &10^{-5}\left(\frac{F_{\rm LW0}}{2\times
			      10^{-17}{\rm
			      erg\;s^{-1}\;cm^{-2}\;Hz^{-1}}}\right)^{-4}\nonumber\\
&\times & \left(\frac{n_{\rm N}}{10^{4}{\rm
			      cm^{-3}}}\right)^{7/2}\left(\frac{T}{10^3{\rm
			      K}}\right)^{11/2} \label{eq:eqh2_2}
\end{eqnarray}
Thus we have obtained the formula to assess H$_2$ fraction with given LW
flux, density and temperature\footnote{Remark that equation
(\ref{eq:eqh2_2}) is valid for $N_{\rm H_2}>10^{14}{\rm cm^2}$.}
.
It is worth noting that $y_{\rm H_2}$ is quite sensitive to density and
temperature. If we consider the adiabatic collapse case, 
the relation $y_{\rm H_2} \propto T^{43/4} = T^{10.75}$ is satisfied,  
because of the adiabatic relation $n_{\rm N} \propto T^{3/2}$. 
We have already found this delicate behaviour of $y_{\rm H_2}$ in the
previous section, where the very quick recovery of H$_2$ fraction is
found in the adiabatic collapse phase (Figure \ref{fig:h2center}).
Remark that this strong dependence on temperature/density holds only if
the H$_2$ is in chemical equilibrium. The relation is not correct for
the very dense regions in which the LW shielding is significant and the
dissociation time scale is much longer than the free-fall time. However,
H$_2$ fraction is large enough to cool the gas cloud in such dense
regions even without such delicate behaviour.

\subsection{Cooling criteria}
Now we are ready to find the cooling criteria. We can evaluate the
cooling time utilizing the equation (\ref{eq:eqh2_2}). The cooling
function around $10^3$K is approximated as 
\begin{equation}
\Lambda_{\rm H_2} \simeq \left\{
\begin{array}{cc}
4.0\times 10^{-25} n_{\rm N}^2 y_{\rm H_2}T_3^4,
~~~~n_{\rm N}\ll 10^4 {\rm cm^{-3}} &\\
~~~&\\
2.5\times 10^{-21} n_{\rm N} y_{\rm H_2}T_3^{4.7},
~~~n_{\rm N}\gg 10^4 {\rm cm^{-3}} &
\end{array}
\right.\label{eq:cool}
\end{equation}
Here the unit of $\Lambda_{\rm H_2}$ is ${\rm erg\;cm^{-3}s^{-1}}$, and $T_3 $ denotes $T/10^3$K.
This formula is obtained by fitting the results of high and low
density limit in \citet{GP98}. 
Then, we can write explicitly the collapse criteria 
$t_{\rm cool} < t_{\rm ff}$  as
\begin{equation}
\frac{3 n_{\rm N} k_{\rm B} T }{2 \Lambda_{\rm H_2}(n_{\rm N}, T,
 F_{\rm LW})} <  \sqrt{\frac{3\pi}{32Gm_{\rm p}n_{\rm N}}}\label{eq:tff_tcool}
\end{equation}
Note that $\Lambda_{\rm H_2}$ is a function of $F_{\rm LW}$ through
$y_{\rm H_2}$.

Combining equations (\ref{eq:eqh2_2}),(\ref{eq:cool}) and
(\ref{eq:tff_tcool}) we obtain the cooling criteria as follows:

\begin{equation}
T  > \left\{
\begin{array}{cc}
7.4\times 10^2 {\rm K} \left(\displaystyle\frac{n_{\rm N}}{10^4 {\rm
	      cm^{-3}}}\right)^{-0.47}\left(\displaystyle\frac{F_{\rm LW0}}{2\times
	      10^{-17} {\rm cgs}}\right)^{0.47} , &\\
~~~~~~~~~~~~~~~~~~~~~~~~~~~~~&\\
~~~~~~~~~~~~~~~~~~~~~{\rm for}~n_{\rm N}\ll 10^4 {\rm cm^{-3}} &\\
~~~&\\
7.9\times 10^2 {\rm K} \left(\displaystyle\frac{n_{\rm N}}{10^4 {\rm
	      cm^{-3}}}\right)^{-0.32}\left(\displaystyle\frac{F_{\rm LW0}}{2\times
	      10^{-17} {\rm cgs}}\right)^{0.43} , &\\
~~~~~~~~~~~~~~~~~~~~~~~~~~~~~&\\
~~~~~~~~~~~~~~~~~~~~~{\rm for}~n_{\rm N}\gg 10^4{\rm cm^{-3}} &
\end{array}
\right.\label{eq:final}
\end{equation}

Therefore, once the temperature at the center of the cloud satisfies above
condition, the cloud cools and collapses even in the presence of
photodissociative radiation field, irrespective of its history on
density-temperature plane.

\subsection{Comparison of the numerical results to the analytic evaluation}
Now the cooling conditions are shown in Figures \ref{fig:nT20} and \ref{fig:nT80} for two cases on
density-temperature plane. Figure \ref{fig:nT20} corresponds to the case
in which a $120M_\odot$ POPIII star is located 20 pc distant from 
the center of the collapsing core, 
whereas it is 80 pc in Figure \ref{fig:nT80}. 
The horizontal axes denote the density at the center of the collapsing
core, vertical axes are the temperature. The upper right hatched regions
satisfy the cooling condition. 
The cooling condition $t_{\rm cool} < t_{\rm ff}$ ( corresponds to the
hatched region) is obtained by directly solving equation (\ref{eq:tff_tcool}).
Note that the whole cooling region and its boundary satisfy
$N_{\rm H_2}>10^{14}{\rm cm^{2}}$, which is required to derive equation
(\ref{eq:eqh2_2}). Thus, equation (\ref{eq:tff_tcool}) based on equation
(\ref{eq:eqh2_2}) correctly describes cooling condition.

The dashed curves starting from the left
side denote the evolutionary tracks on central density-temperature plane,
with no feedback effects for various initial temperatures.
. 
The solid lines which depart from the no feedback
curves represent the cases where the photodissociation feedback is taken
into consideration. In fact, the departing points from the dashed curves
correspond to the timing when the nearby POPIII star is ignited.

We have two distinct classes of runs. 
First one is the case the LW band photons are
effectively shielded. Thus, the cloud core is hardly affected by the
feedback effect. Most of the runs in Figure \ref{fig:nT80} correspond
to this case.

On the other hand, we also have the runs in which the LW
band photons are not shield enough at the onset of the radiative
feedback, which is immediately followed by the adiabatic collapse 
($n_{\rm N} \propto T^{3/2}$).
After this adiabatic phase, we also have two
distinct classes of runs, one of which bounces, and the other
collapses.
The fate of the cloud is obviously determined by the cooling
condition which we derived in this section: if the central density and
temperature of the collapsing cloud can
reach the cooling region, the collapse continues, otherwise it bounces.
In other words, in case the cloud can shrink by the inertia until
the cloud satisfies the cooling condition, the cloud can keep
collapsing. If not, the collapse is stopped by the enhanced thermal
pressure during the adiabatic contraction.

This results basically hold for both of the figures \ref{fig:nT20} and 
\ref{fig:nT80}, in which different
flux is assumed. It is also worth noting that the cooling criteria is
independent of the initial cloud density and temperature, since it is
written by some combination of various reaction rates and physical
constants. In fact, all the calculations starting from different initial
conditions evolve following the cooling criteria.

\subsection{Comparison with the condition $t_{\rm dis} > t_{\rm ff}$}
Now we compare the present results with simple criteria suggested by
\citet{GB01}, i.e.  $t_{\rm dis} > t_{\rm ff}$. However, it is difficult
to compare our results directly 
with theirs since they assume fixed uniform cloud density and H$_2$
fraction. Therefore, we re-derive the condition for given 
$n_{\rm on}$ and $D$
which is equivalent to $t_{\rm dis} < t_{\rm ff}$, utilizing the
relations found in the previous section.

Critical condition $t_{\rm dis} = t_{\rm ff}$ is obtained from the
assumption that H$_2$ is in equilibrium:
\begin{equation}
k_{\rm 2step}y_{\rm H_2} = n_{\rm N}y_{\rm e}k_{\rm H^-} \label{eq:eqh22}
\end{equation}
and the dissociation time equals to the free fall time:
\begin{equation}
k_{\rm 2step}^{-1} = \sqrt{\frac{3\pi}{32G m_{\rm p}n_{\rm N}}}. \label{eq:disff}
\end{equation}
Multiplying the two equations (\ref{eq:eqh22}) (\ref{eq:disff}), and
using equation (\ref{eq:ye_app}), we obtain
\begin{equation}
y_{\rm H_2}=\frac{k_{\rm H^-}}{k_{\rm rec}}.\label{eq:disffyh2}
\end{equation}

Combining equations (\ref{eq:disffyh2}),(\ref{eq:dis}) and
(\ref{eq:Nh2}),
critical distance that satisfy $t_{\rm dis}=t_{\rm ff}$ for a given
core density $n_{\rm N}$ and temperature $T$ is described as 
\begin{equation}
D_{\rm cr}=113{\rm pc}\left(\frac{L_{\rm LW}}{10^{24}{\rm
	    erg/s}}\right)^{-\frac{1}{2}}\left(\frac{n_{\rm N}}{10^3{\rm
	    cm^{-3}}}\right)^{-\frac{7}{16}}\left(\frac{T}{600{\rm K}}\right)^{-\frac{3}{4}}.\label{eq:dcrit}
\end{equation}

Here we use the usual relation between the luminosity of the
star($L_{\rm LW}$) and unshielded flux ($F_{\rm LW0}$),
\begin{equation}
4\pi F_{\rm LW0}D_{\rm cr}^2=L_{\rm LW}
\end{equation}

Figure \ref{fig:LW100} shows the loci on $n_{\rm on} - D$ plane along
which $t_{\rm dis} = t_{\rm ff}$ is satisfied (i.e. $D=D_{\rm cr}$) at the ignition time of
the source star. Two lines correspond to the core temperatures (in
equation \ref{eq:dcrit}) equal to
$300$K and $400$K. If we consider $T_{\rm ini}=100$K case, the core
temperatures at the turn-on time of the source star roughly satisfy
$300{\rm K}\la T \la 400$K (see Figure \ref{fig:norad}). Thus, the
actual condition $t_{\rm dis}=t_{\rm ff}$ would be between the two lines.
Numerical results with $T_{\rm ini}=100$K are also superimposed on the
$n_{\rm on}-D$ plane as the Figure \ref{fig:all}.

Comparing the condition $t_{\rm dis} = t_{\rm ff}$ with the numerical
results, we find that 1) the simple condition $t_{\rm dis} = t_{\rm ff}$
is not so bad, 2) but it overestimates the critical distance by a factor
of 2-3. For comparison, we also plotted the distances that are factor $C$
times smaller than $D_{\rm cr}$ with $T=400$K. Five lines correspond to
$C=0.1,0.2,0.3,0.4,0.5$, respectively. We find $C=0.3-0.4$ fit the numerical results
better than the original condition $D=D_{\rm cr}$ does.

The disagreements between the condition $t_{\rm dis}=t_{\rm ff}$ and the
numerical results basically comes from the dynamical effects of the
collapsing gas. As discussed in the previous subsections, H$_2$ fraction
in the collapsing core recovers very rapidly following the equation
(\ref{eq:eqh2_2}) during the adiabatic compression phase. As a result,
dynamically contracting clouds can keep collapsing even if condition $t_{\rm
dis}< t_{\rm ff}$ is satisfied when the source star is turned-on.

Figure \ref{fig:LW350} also shows the comparison of numerical results
for $T_{\rm ini}=350$K with the criteria $t_{\rm dis}=t_{\rm ff}$. Since
the core temperature at the turn-on time is approximately $100-200$K for
this case, we
plot the loci $t_{\rm dis}=t_{\rm ff}$ for $T=100$K and $200$K.
In this case, we also find qualitatively the same results as above,
however, 
the disagreement is slightly smaller than the previous case. In fact,
we also plotted the loci along which $D=C D_{\rm cr}(200{\rm K})$ are
satisfied for various $C$. In this case, $C=0.5-0.6$ gives a better fit
to the numerical results.
The smaller
difference is due to the fact that the core temperature is so low at
the turn-on time, that the adiabatic collapse is not enough to drive
abundant H$_2$ formation. 

In conclusion, $t_{\rm dis} > t_{\rm ff}$(i.e. $D > D_{\rm cr}$) 
gives a rough estimates, but
we cannot neglect the dynamical effects on the H$_2$ formation.

\section{Discussion}
\label{discussion}
Present calculations are based upon the assumption that the collapsing
gas is self-gravitating. This assumption would be correct for
the prestellar cores formed by some fragmentation process of the shocked
gas in large halos with $T_{\rm vir} \ga 10^4$K. It is also
relevant for the the collapsing 
dense prestellar cores ($n_{\rm N}\ga 10^3{\rm
cm^{^-3}}$) at the center of the very first halos formed directly from CDM
density fluctuations. 
However, it would
not be correct for the earlier phase (i.e. low density phase) of such
prestellar cores with $n_{\rm N}\la 10^3 {\rm cm^{-3}}$, since dark
matter halos are dominat source of gravity at such early phase.
Thus, we have to consider
the effect of dark matter for such low density cores. \citet{AS06} have
performed one dimensional radiation hydrodynamics simulations on this
issue. They take into account the effects by photoionization as well
as the LW radiation. They also consider the effects of the age of POPIII
stars. As a result, they find that net effect of radiation by a nearby
star is approximately even. i.e. negative feedbacks balance with the
positive feedbacks. 
These results could also be checked by our code in future.

LW-band radiation is basically the sum of line emissions, which should
be treated carefully. Unfortunately, it is almost impossible to solve line
transfer problem directly in huge three dimensional simulations by present
computational resources. Therefore, we have employed the self-shielding
function introduced by \citet{DB96}. However, their formula is basically
derived by the calculation for static slab, which might be irrelevant to
the present issue. It is obvious that the self-shielding function is
inappropriate to represent the absorption of the radiation in case the
motion of the fluid element is supersonic. Therefore,
the absorption by the gas in the envelope of the collapsing cloud would
be over estimated, since the infalling envelope is supersonic\citep{Larson69}. However, if the gas motion is not supersonic, the
function gives a good estimate for the absorption of LW photons. In fact,
in our numerical simulations,
most of the absorption takes place in the collapsing core which is
subsonic. Thus, we guess that we evaluate the absorption of LW radiation properly.

The effects of ionizing photons are obviously quite important, although it
is not included in the present paper. Photoheating effects followed by
the photoionization can leads negative feedback effects such as 
the photoevaporation of the cloud (e.g. Yoshida et al. 2006),
creation of shock front (M-type I-front) 
which blow out the whole cloud\citep{SU06}, whereas it also has positive
feedback effects such as enhanced H$_2$ formation\citep{AS06}, or
formation of ``H$_2$ barrier'' between the source and the collapsing
cloud\citep{SU06}. 
This is a quite complicated problem, however, we also have performed
numerical simulations on this issue.
The details will be discussed in the forthcoming paper.

\section{Summary}
We perform radiation hydrodynamics simulations on the radiative feedback
of POPIII stars. We obtain the well-understood collapse criteria of the primordial
prestellar core both analytically and
numerically in the presence of Lyman-Werner band photons from nearby stars. 
The criteria dictate the importance of self-shielding
effects coupled with hydrodynamics. Consequently, the LW radiation
from a POPIII star cannot halt the collapse of the dense clump($n_{\rm
N} \ga 10^3-10^4 {\rm cm^{-3}}$) even if they are in the same halo. 
We also evaluate the effects of diffuse LW radiation
background, which also is not important for the collapse of dense cores
with $n_{\rm N} \ga 10^3{\rm cm^{-3}}$.

\bigskip
We thank the anonymous referee for very careful reading and comments.
We also thank N. Yoshida and M.Umemura for discussions and careful reading
of the manuscript.
N. Shibazaki and K. Ohsuga are acknowledged for continuous encouragements. 
The analysis has been made with computational facilities 
at Center for Computational Science in University of Tsukuba and Rikkyo
University. 
This work was supported in part by Ministry of Education, Culture,
Sports, Science, and Technology (MEXT), Grants-in-Aid, Specially
Promoted Research 16002003 and Young Scientists (B) 17740110.




\setcounter{figure}{0}

\clearpage

\begin{figure}[ht]
\begin{center}
\includegraphics[angle=0,width=10cm]{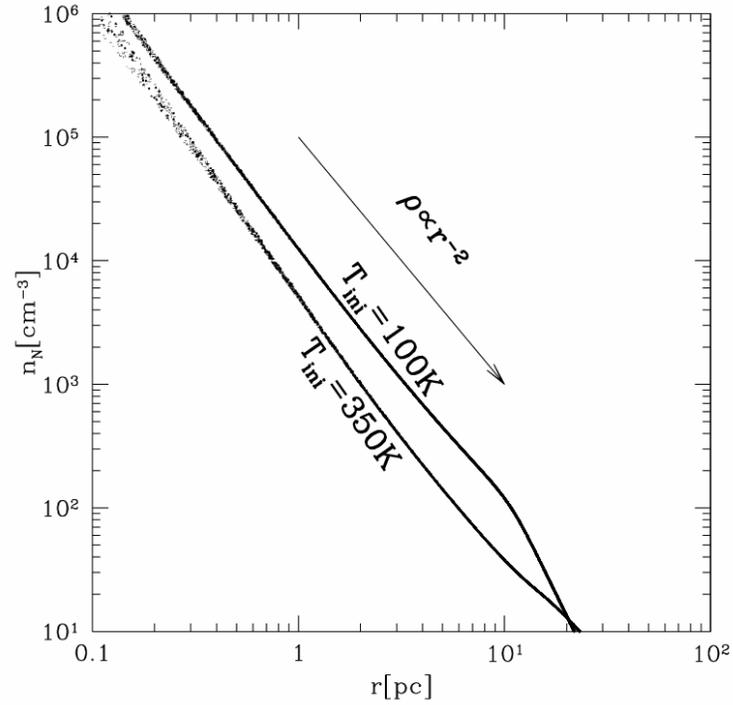}
\caption{Density distribution of the collapsing cloud without radiative
 feedback. Two groups of dots represents the runs with $T_{\rm ini}=100$K
 and $350$K, at the epoch when the central part go below the resolution limit.
}\label{fig:rhor_norad}
\end{center}
\end{figure}

\begin{figure}[ht]
\begin{center}
\includegraphics[angle=0,width=10cm]{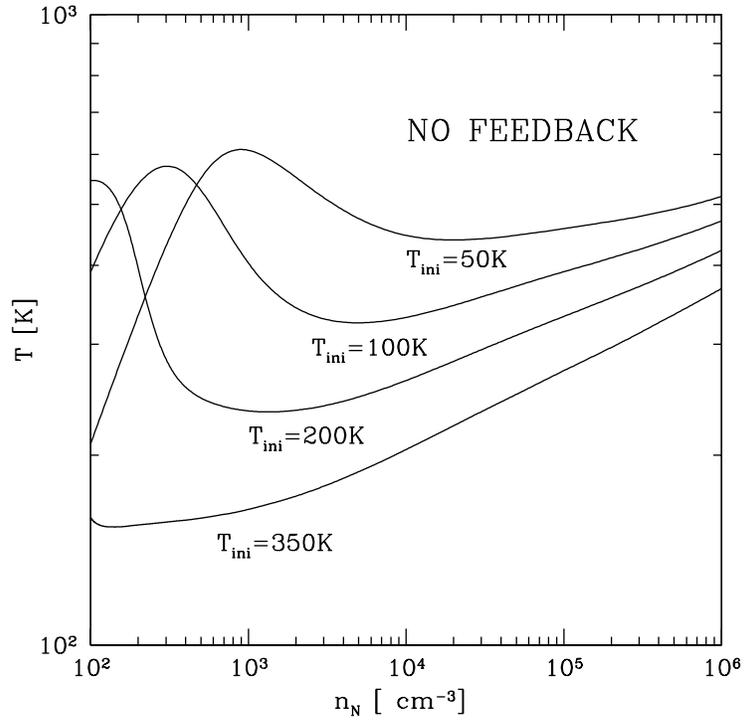}
\caption{Evolution of the central core on density-temperature plane for
 various initial conditions. In these runs, the radiative feedback
 effects are not included.
}\label{fig:norad}
\end{center}
\end{figure}

\begin{figure}[ht]
\begin{center}
\includegraphics[angle=0,width=11cm]{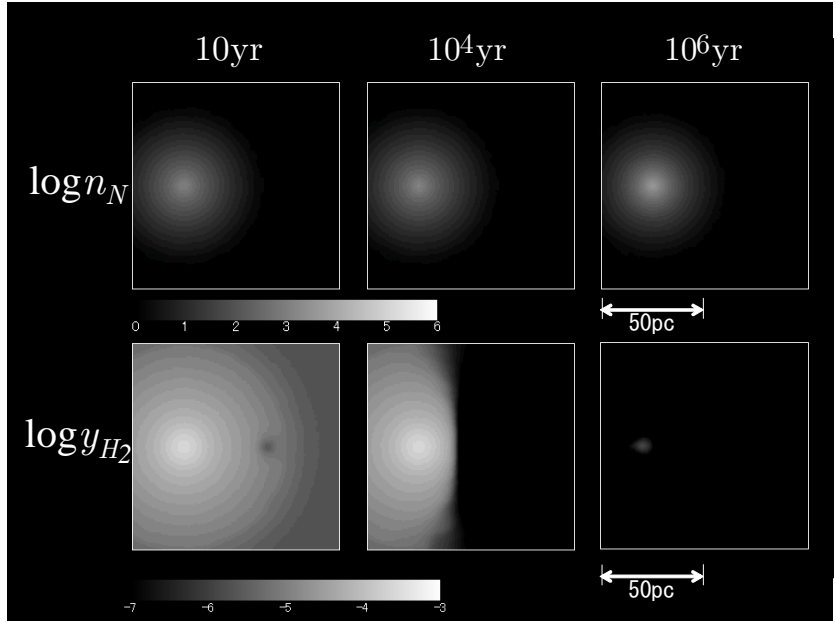}
\caption{Time evolution of the density distribution and H$_2$
 distribution are shown for $D=40{\rm pc}$, $n_{\rm on}=10^3 {\rm cm^{-3}}$,
 $T_{\rm ini} = 100{\rm K}$. Top
 panels represent the 
 density distribution on the slice which include the axis of
 symmetry after the ignition of nearby star. Three panels correspond to
 $t=10$yr, $10^4$yr and $10^6$yr, respectively. Bottom panels also show
 the snapshots of $y_{\rm H_2}$ distribution at three epochs.
}\label{fig:40pc}
\end{center}
\end{figure}

\begin{figure}[ht]
\begin{center}
\includegraphics[angle=0,width=11cm]{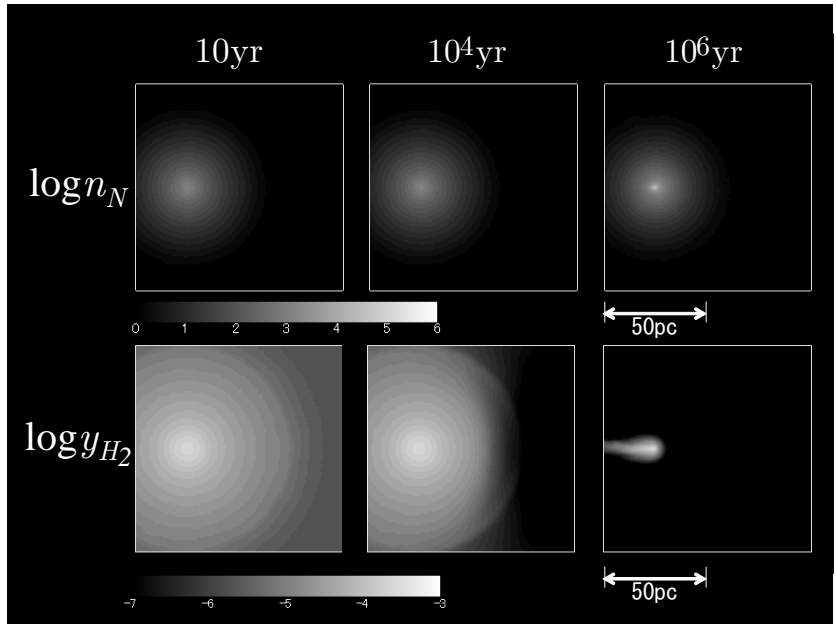}
\caption{
 Same as Figure \ref{fig:40pc} except $D=100{\rm pc}$. 
}\label{fig:100pc}
\end{center}
\end{figure}

\begin{figure}[ht]
\begin{center}
\includegraphics[angle=0,width=10cm]{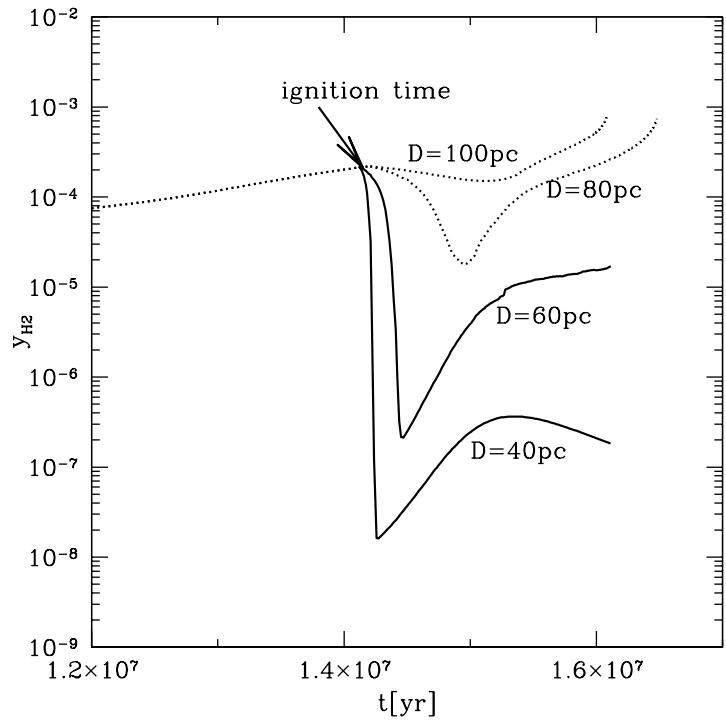}
\caption{
H$_2$ fraction at the center of the cloud is shown as functions of
 time. Results from four runs are shown, in which $n_{\rm on}=10^3{\rm
 cm^{-3}},~ T_{\rm ini}=100{\rm K}$, and $D=40,60,80,100$pc,
 respectively. The horizontal axis denotes the time after the collapse
 of the cloud is started. The ignition time of the nearby source star is
 marked by the arrow. Two solid line represent the case in which the
 cloud fails to collapse ($D=40$pc) and delayed collapse($D=60$pc),
 whereas the successful collapse cases are shown by the two dotted lines($D=80,100$pc). 
}\label{fig:h2center}
\end{center}
\end{figure}

\begin{figure}[ht]
\begin{center}
\includegraphics[angle=0,width=14cm]{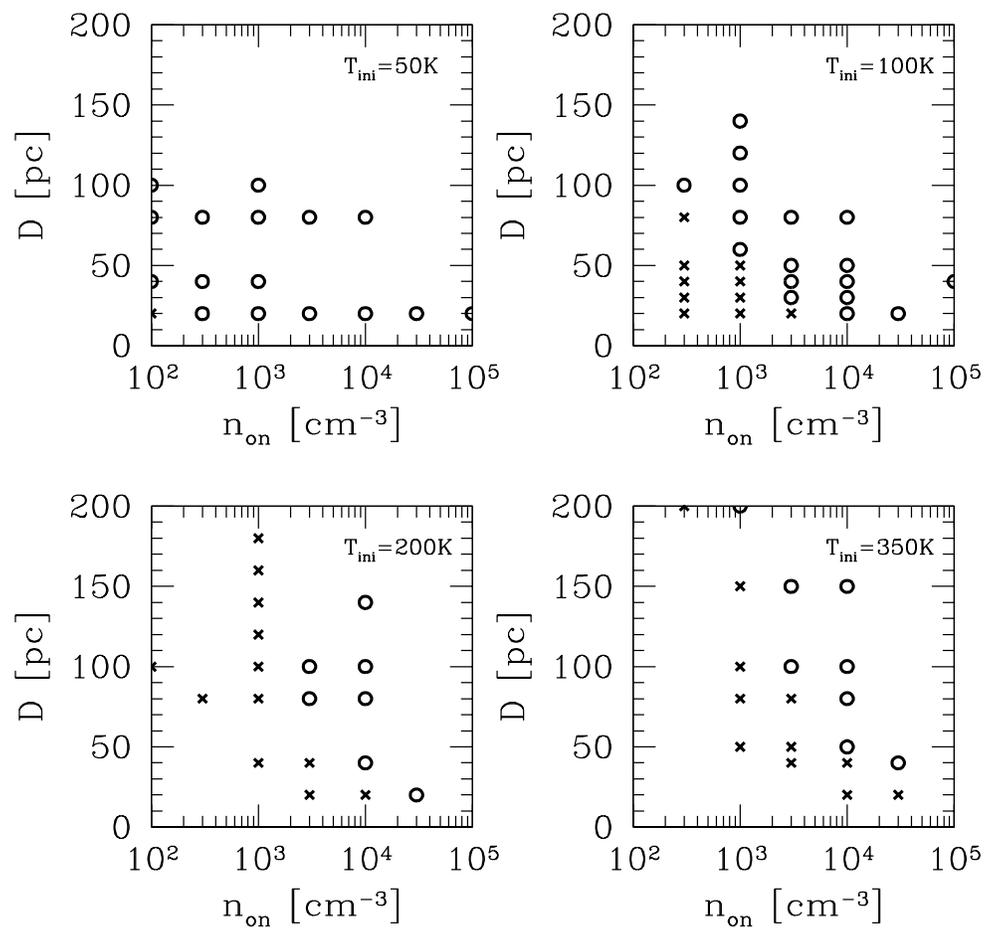}
\caption{Summary of numerical runs are shown on $n_{\rm on} - D$
 plane. Four panels correspond to the runs with $T_{\rm ini} = 50$K ,
 100K, 200K and 350K, respectively. 
Open circles denote the runs in which the clouds collapse successfully, whereas vertices represents the failed collapse.}\label{fig:all}
\end{center}
\end{figure}

\begin{figure}[ht]
\begin{center}
\includegraphics[angle=0,width=10cm]{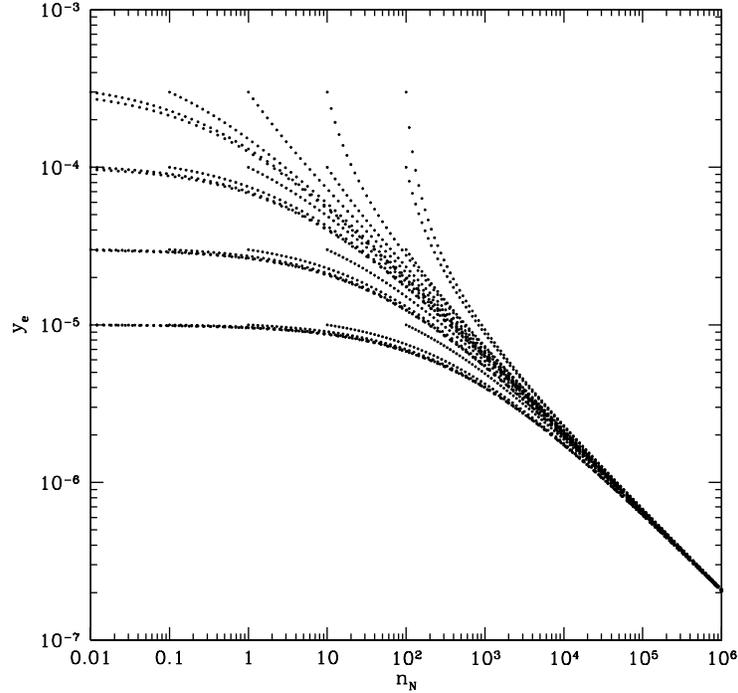}
\caption{Evolutionary tracks of electron fraction $y_{\rm e}$ are plotted
 against number density of the collapsing cloud. Each curve corresponds
 to a certain initial condition $(n_{\rm N0}, y_{\rm e0})$.    
}\label{fig:ye}
\end{center}
\end{figure}

\begin{figure}
\begin{center}
\includegraphics[angle=0,width=10cm]{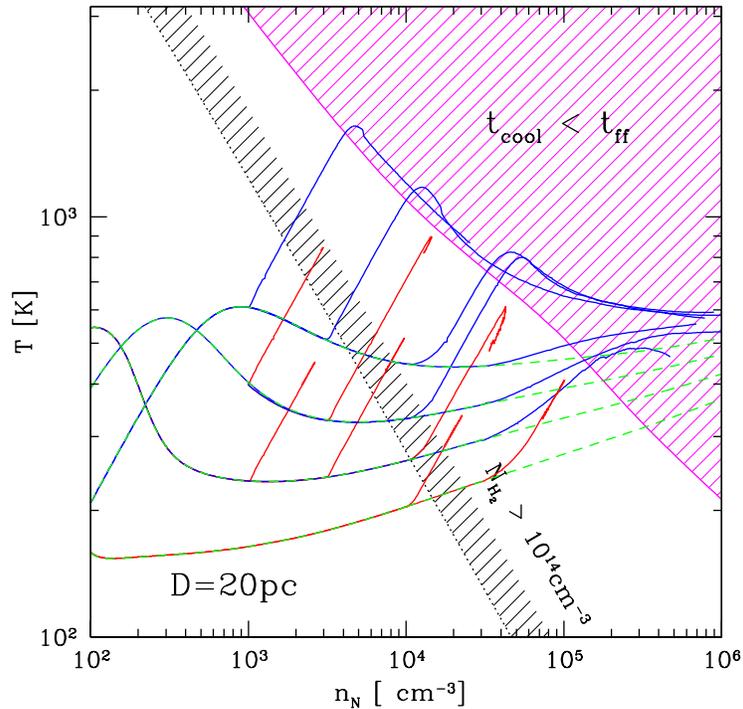}
\caption{Evolutionary tracks on the central density v.s. temperature
 plane of the collapsing core are shown. 
 Horizontal axis denotes the density of the core, whereas
 the vertical axis denotes the temperature. Dashed lines
 are the results of runs without radiative feedback, staring from
 four different initial conditions.
 Solid lines denote the case the results with radiative feedback for various ignition densities $n_{\rm on}$ and $D=20$pc, 
 which in turn collapse successfully or fail to collapse.
 In the upper right shaded region,
 the condition $t_{\rm ff} > t_{\rm cool }$ is satisfied, whereas the  
 upper right region bounded by thin dotted line shows the condition 
where the H$2$ column density of the core is larger than $10^{14}{\rm
 cm^{-2}}$, assuming equation (\ref{eq:eqh2_2}).
 }\label{fig:nT20}
\end{center}
\end{figure}
\begin{figure}
\begin{center}
\includegraphics[angle=0,width=10cm]{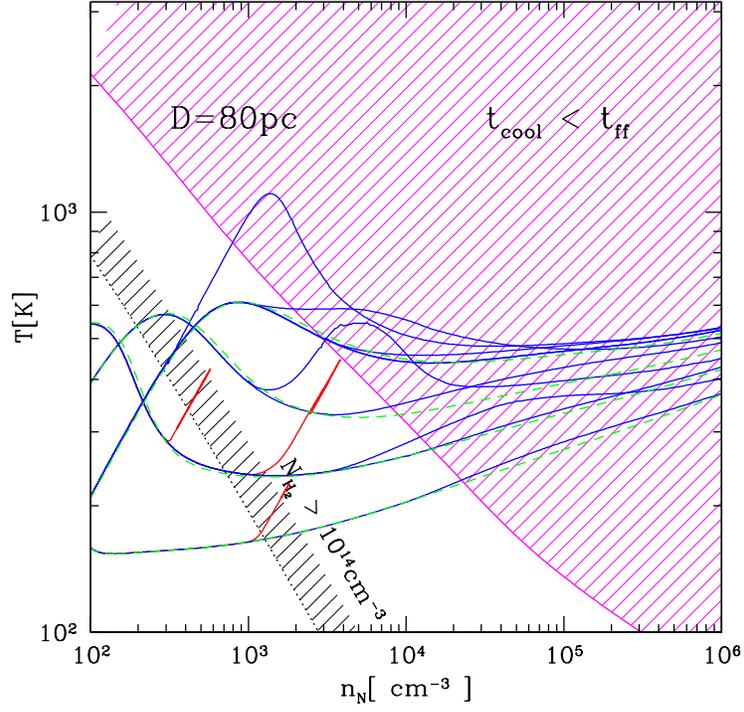}
\caption{Same as Fig. \ref{fig:nT20}, except that $D=80$pc.}\label{fig:nT80}
\end{center}
\end{figure}

\begin{figure}[ht]
\begin{center}
\includegraphics[angle=0,width=10cm]{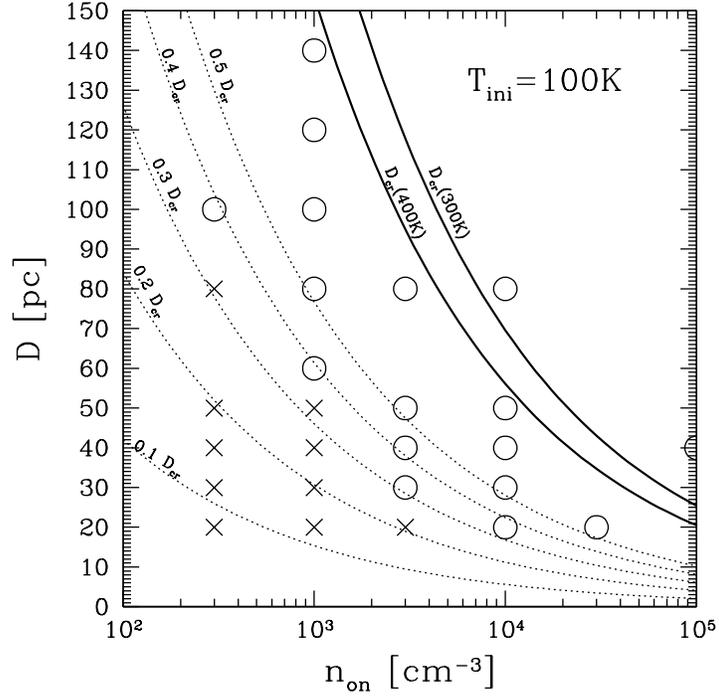}
\caption{The conditions $t_{\rm dis}=t_{\rm ff}$($D=D_{\rm cr}$) are plotted by solid
 lines on $n_{\rm
 on} - D$ plane for $T=300,400$K.
Symbols represent the numerical runs. 
Open circles denote the runs in which the clouds collapse
successfully, whereas vertices represents the failed collapse.
The dotted lines represents the loci along which $D=C D_{\rm cr}(400K)$
 are satisfied. $C=0.1,0.2,0.3,0.4,0.5$ corresponds to each line. 
}\label{fig:LW100}
\end{center}
\end{figure}
\begin{figure}[ht]
\begin{center}
\includegraphics[angle=0,width=10cm]{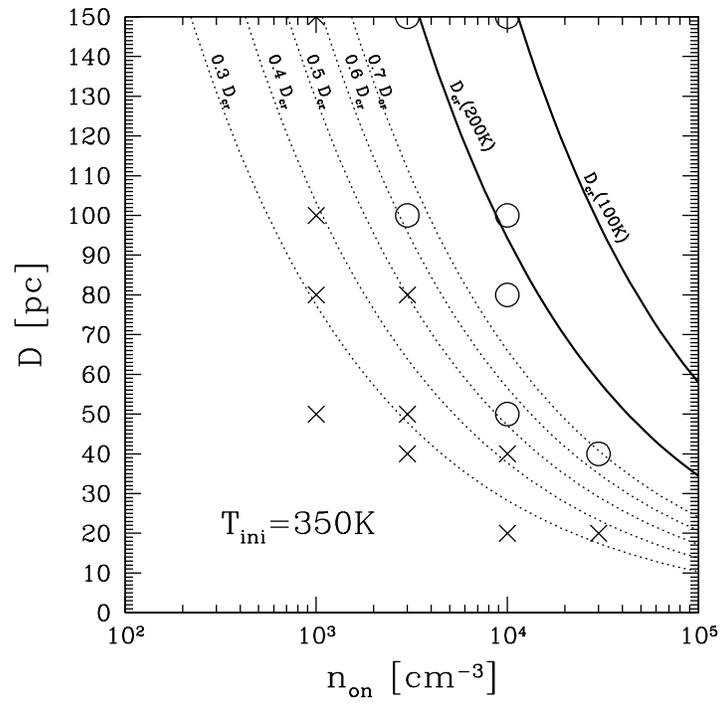}
\caption{Same as Figure\ref{fig:LW100}, except that $T_{\rm ini}=350$K
 and the assumed core temperatures are $100$K and $200$K.
The dotted lines represents the loci along which $D=C D_{\rm cr}(100K)$
 are satisfied. $C=0.3,0.4,0.5,0.6,0.7$ corresponds to each line. 
}\label{fig:LW350}
\end{center}
\end{figure}

\end{document}